\documentclass[preprint,prd,amsmath,amssymb,nofootinbib,tightenlines,floatfix]{revtex4}

\newcommand{\lsim}{~{}_{\textstyle\sim}^{\textstyle <}~}

\newcommand{\be}{\begin{equation}}
\newcommand{\ee}{\end{equation}}

\newcommand{\bea}{\begin{eqnarray}}
\newcommand{\eea}{\end{eqnarray}}

\begin{document}

\preprint{Caltech MAP-318}

\title{\Large Probing the Fundamental Symmetries of the Early Universe: The Low Energy Frontier}


\author{M. J. Ramsey-Musolf}
\affiliation{California Institute of Technology, Pasadena, CA 91125, USA}

\begin{abstract}
Searching for the fundamental symmetries that characterize the particle physics of the early universe lies at the forefront of particle physics, nuclear physics, and cosmology. In this talk, I review low energy probes of these symmetries and discuss what they may teach us about what lies beyond the fundamental symmetries of the Standard Model. 

\end{abstract}

\maketitle



\section{Introduction}

A key quest lying at the intersection of particle physics, nuclear physics, and cosmology is to explain how the properties and interactions of elementary particles in the early universe  shaped the universe as we know it today. Tremendous progress has been achieved in describing fundamental interactions by exploiting the ideas of symmetry, culminating in the relatively simple and remarkably successful Standard Model of electroweak and strong interactions. We know, however, that the  Standard Model (SM) and its symmetries leave open a number of questions that pertain to both the nature of the forces of nature and their role in the dynamics of the unfolding cosmos. Thus, the effort to find out what lies beyond the SM is central to the future of this arena of intersections between particle/nuclear physics and cosmology. 

Two frontiers define the search for new physics: the \lq\lq energy frontier" involving high energy collider studies at Fermilab, the Large Hadron Collider (LHC), and International Linear Collider; and the \lq\lq precision frontier", involving exquisitely precise measurements of electroweak observables carried out primarily (but not exclusively) at low energies. In this talk I will focus on the low energy precision frontier, as others in this meeting have described what lies on the horizon at the energy frontier. In doing so, I will try to address two questions:

\begin{itemize}

\item[i)] {\em What were the fundamental symmetries that governed the microphysics of the early universe? }

\item[ii)] {\em What insights can precision low energy ($E \ll M_Z$) electroweak studies provide into the first question?}

\end{itemize}

In thinking about the microphysics of the universe and its history, it is useful to break the cosmic timeline into two broad periods: one that I call the era of SM success, starting roughly at the time of electroweak symmetry-breaking ($T_{\rm wk} \sim 100$ GeV) and lasting to the present; and the other -- the era of \lq\lq SM puzzles" -- that preceeds it going back to the end of inflation. Most of what we know about the structure and interactions of atoms and nuclei, the abundances of the elements, and dynamics of stars can be described at a fundamental level using the ingredients of the SM. Of course, there remains considerable \lq\lq unfinished business" for the SM in providing detailed, quantitative accounts for phenomena like the QCD phase transition, the large singlet scattering length associated with the small deuteron binding energy, and the dynamics of supernovae. But by and large, we are quite comfortable with the SM as the context in which to address these issues. 

Looking to the time before electroweak symmetry breaking, however, we encounter questions of a for which there exist no satisfying SM answers. Consider three: 

\begin{itemize}

\item[(1)] How was gravity related to the electroweak and strong forces in the beginning? Was there unification of all forces? If all we have are the symmetries of the SM, then the electroweak and strong couplings do not quite unify at high scales. However, they point in the direction of unification, and new symmetries -- such as supersymmetry (SUSY) or extended gauge symmetries -- can do the trick.

\item[(2)] The Planck scale associated with \lq\lq strong" gravity, $M_{\rm P}\sim 2\times 10^{18}$ GeV, is sixteen orders of magnitude larger than the scale of electroweak symmetry breaking, $M_{\rm wk}$. Why didn't physics at distances of order $1/M_{\rm P}$ push $M_{\rm wk}$ up to a much higher scale via loop effects? To put it another way, why is the Fermi constant that varies as $1/M_{\rm wk}^2$ so large? The existence of new symmetries or additional spacetime dimensions can answer this question.

\item[(3)] The effects of inflation were so dramatic that the initial conditions of the post-inflationary era were presumably as symmetric as they could be in every way we might think of. In this case, how did the universe develop the small excess of baryonic matter over antimatter that is responsible for the existence of galaxies, planets, and human life? The particle physics of the SM falls many orders of magnitude short of explaining the observed baryon excess, so clearly new forces and symmetries are needed.

\end{itemize}

In addition to these questions, we also face new puzzles associated with the neutrino. The existence of its non-zero mass that is many orders of magnitude smaller than that of the other SM particles suggests the violation of two symmetries of the minimal SM -- lepton flavor and total lepton number -- in the dynamics of the early universe. What we don't know is the structure of the larger theory that incorporates these symmetry violations in a natural way. 

In the rest of the talk, then, I will focus on what new clues we can gain from low energy studies about the answers to these questions. In doing so, I will organize the discussion around three questions contained in the recent NSAC subcommittee "Tribble Report" and the associated low-energy precision studies\cite{NSAC05}:

\begin{itemize}

\item[i)] {\em Why is there more matter than anti-matter in the present universe?} Here I will discuss the implications of searches for the electric dipole moments of various particles.

\item[ii)] {\em What are the unseen forces that disappeared from view as the universe cooled?} We know that such forces -- associated with new, fundamental symmetries -- must exist in order to address all of the puzzles discussed above. I will review what studies of weak decays and lepton scattering may teach us about those symmetries.

\item[iii)] {\em What are the masses of neutrinos and how have they shaped the evolution of the universe?} Since others in this meeting have discussed neutrino physics in detail, I will devote less effort to this question than to the other other two. However, I will highlight a few aspects of neutrinos and symmetries in discussing precision electroweak processes.

\end{itemize}

The field of studies that is addressing these questions is remarkably diverse, and I cannot hope to provide a comprehensive review here. Rather, I will highlight a few areas that illustrate the richness of the low energy precision frontier, apologize to those whose work I do not have space to include here, and refer the reader to additional references along the way.

\section{Electric Dipole Moments and Baryogenesis}

If the initial, post-inflationary conditions were matter-antimatter symmetric, then the particle physics of the post-inflationary era would have to be responsible for generating the baryon asymmetry. It is convenient to express the asymmetry in terms of the ratio $Y_B=\rho_B/s_\gamma$, where $\rho_B$ is the net baryon density and $s_\gamma$ is the photon entropy density at freeze-out. The WMAP analysis of the comsic microwave background and Big Bang Nucleosynthesis (BBN) both tell us that $Y_B\sim 10^{-10}$ \cite{Spergel:2003cb,Eidelman:2004wy}. Forty years ago, Sakharov identified the three key ingredients
for any successful accounting for this number\cite{Sakharov:1967dj}: (1) a violation of baryon number ($B$) conservation; (2) a violation of both $C$ and $CP$ symmetries; and (3) a departure from thermal equilibrium at some point during cosmic evolution\footnote{The last ingredient can be replaced by $CPT$ violation, and some baryogenesis scenarios exploit this fact.}. 

What we don't know is where in the post-inflationary epoch these three ingredients came together to generate $Y_B\not= 0$.  One possibility is that baryogenesis took place at very early times, associated with particle physics at scales $\gg M_{\rm wk}$. Another possibility involves baryogenesis at the weak scale during the electroweak phase transition. It could also have occurred at any point between these two extremes. One advantage of considering electroweak baryogenesis (EWB) is that it is the scenario most readily tested by laboratory experiments. In contrast, higher-scale scenarios, such as GUT baryogenesis or leptogenesis are intellectually appealing but harder to test experimentally. At least in the case of leptogenesis we can look for some of the elements that have to exist to make this scenario plausible: CP-violation in the lepton sector, an appropriate scale for $m_\nu$; and lepton number violation. Neutrino oscillation studies -- together with ordinary $\beta$-decay and neutrinoless double $\beta$-decay ($0\nu\beta\beta$) -- will look for these elements. Here, however, I will focus on EWB since we have more detailed probes of this scenario -- such as EDM searches. In principle, the right combination of experimental information could rule EWB out altogether, leaving us with more speculative, higher-scale baryogenesis as our only alternative.

The most obvious context for considering EWB is the SM, as it contains all of Sakharov's criteria. Baryon number violation takes place through the thermal excitation of gauge configurations called sphalerons that can decay to a different electroweak vacuum thereby changing baryon number. In principle, the electroweak phase transition provides the needed departure from thermal equilibrium, shutting off the sphaleron processes and freezing non-zero baryon number into the region of broken electroweak symmetry. However, this transition needs to be strongly first order to prevent thermal fluctuations from \lq\lq washing out" baryon number. Moreover, the extent to which the  transition is first order depends on the parameters of the Higgs potential. Given the experimental lower bounds on the Higgs mass, we know that the SM cannot have provided a strong first order transition. Finally, the SM contains CP-violation via the phase in the CKM matrix that affects interactions between the spacetime varying Higgs vev and quarks at the phase transition boundary. Unfortunately for the SM, the net effect of this CP violation is highly suppressed by the Jarlskog invariant and multiple powers of quark Yukawa couplings\cite{Shaposhnikov:1987tw}. In short, the SM has the right ingredients, but the EWPT is not strongly first order and the effects of CP-violation too weak to explain even the tiny value of $Y_B$.

Various models for new physics at the weak scale can remedy these SM shortcomings\cite{Riotto:1999yt}. In the case of supersymmetric models, for example, the presence of additional scalar degrees of freedom in the Higgs potential can lead to a strong first order EWPT. Similarly, new CP-violating interactions between the Higgs fields and superpartners of the SM particles are not suppressed as in the case of SM CP-violation, so that it is possible to have a sufficient asymmetry between reaction rates for particle and antiparticle creation, leading ultimately to a net baryon asymmetry of the magnitude implied by BBN and the CMB. Whether or not one obtains the observed value of $Y_B$ in a given model depends in detail on the quantum transport dynamics in the plasma at the phase boundary and on the values of the model parameters. Thus, improvements are needed in both theory and experiment. From the latter standpoint, we are on the cusp of a new generation of EDM searches that will improve the sensitivity by anywhere from two- to four-orders of magnitude beyond current limits. While these experiments will not be able to see EDMs associated with SM CKM CP-violation, one has reason to expect non-zero results of new, weak scale CP-violation is responsible for $Y_B$. In this meeting, Dave DeMille and Brad Filippone have discussed the efforts to improve the electron and neutron EDM sensitivity, and there exist many other similar efforts involving the EDMs of neutral atoms, the muon, and even the deuteron (for a recent summary, see Ref.~\cite{Erler:2004cx}). 

Theoretically, a robust confrontation between these experiments and expectations based on the baryon asymmetry requires many refinements of $Y_B$ and EDM calculations. Efforts to sharpen EDM calculations are on-going. In the last few years, however, there has been a new push to put $Y_B$ computations on firmer ground using the techniques of non-equilibrium quantum field theory\cite{riotto}. While I do not have time to go into details here, I believe that his development is an important one to highlight, since -- at the end of the day -- one would like to know how reliably one can constrain EWB via the new EDM measurements. It is also important to note that the phenomenological inputs are not limited to EDM experiments, though they are crucial. We also require knowledge about other aspects of new physics models -- such as the spectrum -- and this knowledge will be provided by both LHC/ILC studies and low energy precision electroweak measurements.

\section{Weak Decays \& Lepton Scattering}

One window on the possibility  new forces and associated symmetries at the weak scale is provided by precise studies of the weak decays of light quarks and leptons and of lepton scattering. The strategy of these studies to look either for tiny deviations from SM predictions for decay rates, asymmetries, {\em etc.}, or for non-vanishing observables whose existence is either forbidden or highly suppressed in the SM. Unlike collider experiments that try to produce new particles associated with the previously \lq\lq unseen" forces that decoupled at earlier times, precision decay and scattering studies look for their footprints in the pattern of deviations from -- or agreement with -- SM predictions. The history of the SM tells us that the pursuit of both directions is important. Prior to the discovery of the top quark at the Tevatron, for example, precise studies of electroweak observables at the level of radiative corrections told us what to expect for its mass. 
The measured value was  consistent with those expectations -- making the discovery of the top quark a triumph for the SM.

Going beyond the SM requires pushing both the precision and energy frontiers. Just as higher energies are needed to make the new particles in the lab, higher precision is required to discern their virtual effects -- since the latter generally become weaker the heavier the new physics scale.
Clearly, there is substantial, on-going progress on both fronts. In the case of precision physics, the recent measurements of the muon anomalous magnetic moment, $a_\mu$, provides an illustration. To the extent that theoretical issues involving SM hadronic vacuum polarization and light-by-light contributions are resolved, the deviation from the SM prediction seems most likely within SUSY at large $\tan\beta$. As I will outline below, the results from other, low-energy precision electroweak studies could provide similar clues about the existence and character of new fundamental symmetries at the electroweak scale. For concreteness, I will concentrate on four examples.

\noindent {\em $\beta$-decay and CKM unitarity.} Tests of the unitarity of the first row of the CKM matrix has long been of interest to many people at this meeting. The most precisely known input in this test is the element $V_{ud}$ that is determined from $\beta$-decay. The other key input is $V_{us}$ that one obtains most reliably from $K_{e3}$ decays. The value of $V_{ub}$ -- though interesting -- is too small to matter for the unitarity test. For many years, there has been a small but irritating deviation from the unitarity requirement, based on these two quantities. The conventional wisdom had been that nuclear structure theory was responsible, since  the most precise value of $V_{ud}$ is obtained from \lq\lq superallowed" nuclear decays. These decays involve transitions between $J^\pi=0^+$ nuclei. In the limit of exact isospin symmetry in the nuclear states, the transition matrix elements are given solely by Clebsch-Gordon coefficients and are independent of nuclear structure details. In real nuclei, of course, isospin symmetry is not exact, and one must compute nuclear structure corrections. Ian Towner and John Hardy have devoted their careers to measuring these transitions and computing the corrections, with the result that the corrected \lq\lq $ft$" values for twelve different superallowed transitions are in agreement to a few parts in $10^4$\cite{Towner:2005qc}. Thus, it seems hard to believe that there are significant nuclear structure corrections not properly taken into account.

Additional theoretical input is required to extract $V_{ud}$ from these $\beta$-decay results. What one obtains from experiment  is a $\beta$-decay Fermi constant (or vector constant), $G_\beta$,  that is related to the Fermi constant determined from the muon lifetime, $G_\mu$ as
\be
G_\beta=G_\mu\, V_{ud}\, \left(1+{\widehat\Delta r}_\beta-{\widehat\Delta r}_\mu\right) \ \ \ ,
\ee
where ${\widehat\Delta r}_{\beta,\mu}$ are ${\cal O}(\alpha/4\pi)$  electroweak radiative corrections to the two decay amplitudes. The quantity ${\widehat\Delta r}_\beta$ contain hadronic structure uncertainties associated with the $W\gamma$ box graphs involving the lepton and nucleon pairs. Recently, Marciano and Sirlin have reduced this uncertainty by relating the short distance part of the one-loop integral to the Bjorken sum rule and by applying large $N_C$ correlators to the pertrubative-nonperturbative transition region\cite{Marciano:2005ec}. The resulting value for $V_{ud}$ obtained by these authors and by an up-dated global analysis of superallowed decays\cite{Hardy:2005qv} is
$ V_{ud}=0.97377(11)(15)(19)$, where the first error is the combined experimental $ft$ error and theoretical nuclear structure uncertianty; the second is associated with nuclear coulomb distortion effects; and the last error is the theoretical hadronic structure uncertainty.

Experimentally, new efforts are underway to extract $G_\beta$ from both neutron decay and $\pi$-decay (for a recent review, see Ref.~\cite{Erler:2004cx}). The latter is the cleanest theoretically, but the experimental precision may not reach the present level obtained from the nuclear decays. New developments using cold and ultracold neutron technology may lead to more progress with the neutron decay. In this case, since the neutron lifetime, $\tau_n$, depends on matrix elements of both the vector and axial vector charged currents, and since the latter receives  a QCD renormalization that one cannot yet reliably compute from first principles, the two must be separated by measuring an observable that has a different dependence on the two components of the current. The most common choice is the asymmetry parameter \lq\lq $A$" associated with the parity-violating correlation between the $e^-$ momentum and neutron spin. Brad Filippone has described the history and progress of these measurements in his contribution, including a discussion of the new LANSCE experiment and efforts at ILL, NIST, and the SNS. 

Despite the intense experimental and theoretical effort to obtain a robust test of first row CKM unitarity, the situation is presently clouded by several developments. First, as discussed extensively elsewhere in this meeting, the value of $V_{us}$ is in flux, as new measurements of $K_{e3}$ branching ratios have shifted the world average. In order to obtain $V_{us}$ from these results, however, one requires a theoretical value of the $K$-to-$\pi$ form factor $f_{+}(t)$ at the photon point, and there exist conflicting theoretical results for this quantity. In the case of $\beta$-decay, new Penning trap measurements of several superallowed $Q$-values\cite{Savard:2005cv} seem to be pointing to a lower value for $V_{ud}$ that would widen the discrepancy with unitarity. On the other hand, a new measurement of $\tau_n$ at ILL shifts the central value by $\sim 6\sigma$\cite{Serebrov:2004zf} and, combined with the latest ILL result for $A$\cite{Abele:2002wc}, would point toward better agreement with unitarity. 

It is important to settle these issues because the implications of the CKM unitarity test for new forces and symmetries could be important. My own recent interest has focused on the implications for SUSY, whose presence can effect ${\widehat\Delta r}_\beta-{\widehat\Delta r}_\mu$ via either one-loop corrections involving superpartners\cite{Kurylov:2001zx} or new tree-level interactions\cite{Ramsey-Musolf:2000qn}. The latter arise if a symmetry called \lq\lq R-parity" is violated -- something that is undesirable from the standpoint of cosmology since it would allow the lightest neutralino to decay to rapidly and prevent its being a viable cold dark matter candidate\footnote{This can be done in a way that still avoids proton decay.}. On the other hand, if R-parity is conserved, then SUSY only contributes via loop effects. In either case, the implications of CKM non-unitarity would be problematic. In the R-parity conserving case, one would have to have a spectrum of superparnters that conflicts with most models for how SUSY is broken at some high scale. If R-parity is conserved, such a conflict can be avoided, but one sacrifices a viable SUSY dark matter candidate. Clearly, one would like to know whether one is forced into either alternative, so the unresolved CKM unitarity situation needs resolution. 

\noindent {\em Parity-violating electron scattering.} An alternate way to probe for the virtual effects of SUSY (with or without R-parity) or other new forces is to measure the parity-violating asymmetries associated with the elastic scattering of longitudinally polarized electrons from fixed targets. These asymmetries have the generic form for a target $f$:
\be
A_{PV} = \frac{G_\mu Q^2}{4\sqrt{2}\pi\alpha}\, \left[Q_W^f+F(Q^2,\nu)\right]
\ee
where $Q^2$ is the momentum transfer-squared, $\nu$ is the dimensionless energy transfer in the lab frame, $Q_W^f$ is the \lq\lq weak charge" of the target, and $F(Q^2,\nu)$ is a form factor term. To probe for new forces and symmetries, one would like to extract $Q_W^f$ since it can be computed within the SM and compared with experiment. Generically, it has the form
\be
Q_W^f={\hat\rho}_{NC}\left[2T_3^f-4Q_f{\hat\kappa}\sin^2{\hat\theta}\right]+{\hat\lambda}_f
\ee
where $Q_f$ and $T_3^f$ are the electric charge and weak isospin, respecitvely,  of the target; and ${\hat\rho}_{NC}$ and ${\hat\kappa}$ are process-independent radiative corrections in the $\overline{\rm MS}$ scheme; ${\hat\lambda}_f$ is a process-dependent radiative correction; and ${\hat\theta}$ is the $\overline{\rm MS}$  running weak mixing angle. 

Two cases -- $f=e$ and $f=p$ -- are of particular interest  since their  weak charges are suppressed: $|Q_W^{e,p}| \lsim 0.1$. This accidental suppression makes them relatively more transparent to the effects of new forces and symmetries, as compared to other unsuppressed SM observables. Moreover, the theoretical uncertainties that enter the analysis of semileptonic decays are far less serious for $Q_W^{e,p}$. One class of uncertainties that enters both weak charges through the running of $\sin^2{\hat\theta}$ involves light-quark contributions to the $Z\gamma$ mixing tensor. Recently, my collaborator J. Erler and I have reanalyzed these contributions and shown that the uncertainty falls well below the level relevant for interpreting the measurements\cite{Erler:2004in}. A second class enters $Q_W^{p}$ only and is associated with QCD effects in the $WW$ and $Z\gamma$ box graphs. As discussed in Refs.~\cite{Marciano:1983ss,Ramsey-Musolf:1999qk,Erler:2003yk}, various features conspire to suppress these uncertainties, leading to a theoretically robust SM prediction. 

Experimentally, $A_{PV}$ has been determined with a {\em tour de force}   measurement by the E158 Collaboration at SLAC
\cite{Anthony:2005pm}, and the results agree nicely with the SM prediction. Plans are being developed for a more precise version of this measurement at Jefferson Lab using the energy-upgraded 12 GeV beam. In the case of $Q_W^p$, a $\sim 4\%$ measurement by the Q-Weak collaboration is planned at the present 6 GeV beam at Jefferson Lab. 
Future measurements for deep-inelastic $eD$ and elastic scattering from $0^+$ nuclei are also being considered (for a recent review, see Ref.~\cite{Erler:2004cx}).

What makes these measurements interesting is their complementary sensitivity to new forces and symmetries. In the case of SUSY, for example, the effect of SUSY loop contributions to ${\hat\rho}_{NC}$, ${\hat\kappa}$, and ${\hat\lambda}_f$ would tend to cause same-sign relative shifts in the values for $Q_W^{e,p}$\cite{Kurylov:2003zh}. In contrast, the presence of R-parity violating interactions could lead to rather sizable, opposite-sign relative changes. A similar complementarity holds for other new forces and symmetries, such as those associated with $U(1)^\prime$ models or leptoquarks\cite{Erler:2003yk}. 

\noindent {\em $\mu$-decay and the Michel spectrum.} Studies of the Michel spectrum and lepton polarization in $\mu$-decay have recently received new attention as a probe of new forces and symmetries. The TWIST Collaboration has carried out new determinations of the Michel parameters $\rho$, $\delta$, and $P_\mu\xi$ at TRIUMF with substantially improved precision \cite{Musser:2004zw,Gaponenko:2004mi}, while the a collaboration at PSI has performed new measurements of the positron polarization with similar advances\cite{Danneberg:2005xv}. A new measurement of the $\mu$ lifetime is also underway at PSI that will reduce the uncertainty in $G_\mu$ by an order of magnitude. Since these experiments have been discussed separately in a dedicated session on muon physics, I will concentrate on some implications for new physics. 

The theoretical analysis of $\mu$-decay studies is conventionally performed with reference to an effective four-fermion Lagrangian
\be
\label{eq:mudecay}
{\cal L}^{\mu-\rm decay} = -\frac{4 G_\mu}{\sqrt{2}}\ \sum_{\gamma,\, \epsilon,\, \mu} \ g^\gamma_{\epsilon\mu}\, 
\ {\bar e}_\epsilon \Gamma^\gamma \nu {\bar\nu} \Gamma_\gamma \mu_\mu
\ee
where the sum runs over Dirac matrices $\Gamma^\gamma= 1$ (S), $\gamma^\alpha$ (V), and $\sigma^{\alpha\beta}/\sqrt{2}$ (T) and the subscripts $\mu$ and $\epsilon$  denote the chirality ($R$, $L$) of the muon and final state lepton, respectively. The Michel parameters depend on different combinations of the effective couplings $g^\gamma_{\epsilon\mu}$. For a recent global analysis, see Ref.~\cite{Gagliardi:2005fg}. Various models have different implications for the $g^\gamma_{\epsilon\mu}$. Loop effects in SUSY, for example, can generate a non-zero $g^S_{RR}$ that contributes to the parameter $\eta$ and that, in turn, will affect the extraction of the Fermi constant from the muon lifetime\cite{mrm06}. If there is a new, low-energy right-handed gauge symmetry as in left-right symmetric models, then one would expect to see a non-vanishing $g^V_{RR}$ or $g^V_{LR,RL}$. Experimental limits on these couplings can lead to constraints on the masses and mixing angles of new right-handed gauge bosons that are competitive with information obtained from other collider and precision electroweak studies.

Recently, it was observed that one may derive model-independent constraints on some of the $g^\gamma_{\epsilon\mu}$ based on the scale of neutrino mass and naturalness considerations\cite{Prezeau:2004md}. In particular, the operators in Eq. (\ref{eq:mudecay}) containing charged leptons of opposite chirality can contribute to $m_\nu$ through loop effects. In the original paper making this observation, the authors derived limits based on two-loop effects. In a follow-up analysis, my collaborators and I observed that one actually obtains stronger constraints from one-loop effects and have carried out a complete, one-loop renormalization group analysis\cite{erwin06}. The most robust constraints apply to the couplings $g^V_{LR, RL}$ and lead to bounds that are at least three orders of magnitude smaller than current experimental limits. For the $g^{S,T}_{LR,RL}$, the situation is more subtle, as there exist gauge-invariant operators that contribute to these couplings that are not constrained by $m_\nu$. Nevertheless, one would not expect the size of their effects to be  significantly larger than those associated with the constrained operators, and in this case the expectation is at least three-orders of magnitude below present experimental sensitivities. Should future $\mu$-decay experiments uncover a significantly larger value for any of these couplings, it would imply non-trivial fine-tuning and/or flavor-structure in models for $m_\nu$.

\noindent {\em Searches for lepton flavor violation.} In the minimal SM, both lepton flavor (LF) and number (LN) are conserved. The observation of neutrino oscillations has told us that LF is not conserved in nature, and it increases the likelihood that LN is also violated by a Majorana neutrino mass term. As discussed by Boris Kayser in this meeting, the experimental search for neutrinoless double $\beta$-decay ($0\nu\beta\beta$) aims to find explicit evidence for LN violation. Assuming a significant non-zero signal is observed\cite{Klapdor-Kleingrothaus:2004wj}, one would also like to use $0\nu\beta\beta$ to determine the absolute scale of $m_\nu$ (at least, for the inverted hierarchy scenario). Apart from the well-known problem of uncertainty in the nuclear matrix elements, this hope faces an additional complication that is generally less highlighted. In many models for new forces at the TeV scale,  LN violating interactions that do not directly involve the light Majorana neutrinos can lead to $0\nu\beta\beta$. Examples include R-parity violating SUSY involving LN violating interactions between sleptons together with exchanges of neutralinos, and  left-right symmetric models wherein the exchanged particle is a heavy Majorana neutrino. If the mass scale associated with these new forces is not too far from the TeV scale, then the contribution to the $0\nu\beta\beta$ rate can be comparable to that from light Majorana neutrino exchange. In this case, one would not know now to extract a value of $m_\nu$ without making model-dependent assumptions. 

Fortunately, there exists at least one experimental diagnostic tool that would help with this problem\cite{Cirigliano:2004tc}. In most models that give rise to LN violation, one also generally predicts non-zero rates for flavor-violation processes involving charged leptons, such as $\mu\to e\gamma$ or $\mu\to e$ conversion in nuclei. Generally speaking, one expects the ratio of the branching ratios for these two processes to be ${\cal O}(\alpha)$ since the conversion process involves one more electromagnetic vertex than the LF-violating decay. However, if the scale of LF violation is relatively light (of order a few TeV), there exist logarithmic enhancements of the conversion operator\cite{Raidal:1997hq} that would render this ratio to be ${\cal O}(1)$. In this case, one would expect the LN violation scale to also be light enough to lead to complications in the interpretation of $0\nu\beta\beta$. On the other hand, if the ratio is ${\cal O}(\alpha)$, then the scale of LN would be too high to be problematic for $0\nu\beta\beta$. 

Experimentally, a new search for $\mu\to e\gamma$ is underway at PSI\cite{Signorelli:2003vw} and, until recently, one hoped for a powerful new search for $\mu\to e$ conversion at the Brookhaven National Laboratory\cite{Popp:2001hu}. Taken together, these two experiments would have provided a powerful diagnostic tool for determining the scale where LF is broken. Unfortunately, the cancellation of the RSVP program at Brookhaven means that an alternate venue for the conversion experiment must be found. It is clearly important that an effort in this direction be pursued. 

\section{Conclusions}

In this discussion, I hope I have given a flavor of the power of precise low energy studies to look probe the fundamental symmetries of the early universe. Clearly, there is considerable excitement in this field, and given the limitations of time and space, many other efforts that I could not review (for a more extensive survey, see Ref.~\cite{Erler:2004cx}). The coming of the LHC will clearly open a new direct window on the possible forces and symmetries at the weak scale and just beyond. In this era, the low energy program will continue to provide information that complements and -- in some cases reaches beyond -- that obtained from colliders. In both cases, we clearly have much to look forward to.


\vspace{-2mm}

\acknowledgments

\vspace{-2mm}
I wish to thank the organizers of the meeting for financial support.
This work was supported in part under U.S. DOE contract
DE-FG02-05ER41361 and NSF grant PHY-0071856.

\end{document}